\documentclass[aps,prb,superscriptaddress,twocolumn,showpacs,amsmath,amssymb,10pt]{revtex4-1}
\usepackage{graphicx}
\usepackage{dcolumn}
\usepackage{bm}
\usepackage{textcomp,gensymb}
\usepackage{amsmath}
\newcommand{\ve}{\boldsymbol}
\newcommand{\tn}{\textnormal}

\newcommand{\D}{\tn{d}} 
\newcommand{\fig}{Fig.~\ref}
\newcommand{\figs}{Figs.~\ref}

\newcommand{\figu}{Figure~\ref}

\newcommand{\eq}{Eq.~\eqref}

\newcommand{\Ref}{Ref.~\onlinecite}

\begin{document}



\title{Ultrafast photocurrents at the surface of the \\ three-dimensional topological insulator $\tn{Bi}_2\tn{Se}_3$}

\author{Lukas~Braun}%
\affiliation{Fritz Haber Institute of the Max Planck Society, 14195 Berlin, Germany}
\author{Gregor~Mussler}
\affiliation{PGI-9 and JARA-FIT, Forschungszentrum J\"ulich, 52425 J\"ulich, Germany}
\author{Andrzej~Hruban}
\affiliation{Institute of Electronic Materials Technology, 01-919 Warsaw, Poland}
\author{Marcin~Konczykowski}
\affiliation{Laboratoire des Solides Irradi\'es, CNRS UMR 7642 \& CEA-DSM-IRAMIS, Ecole
Polytechnique, F91128 Palaiseau, France}
\author{Thomas~Schumann}
\affiliation{Institut f\"ur Physik, Ernst-Moritz-Arndt Universit\"at Greifswald, 17489 Greifswald,
Germany}
\author{Martin~Wolf}
\affiliation{Fritz Haber Institute of the Max Planck Society, 14195 Berlin, Germany}
\author{Markus~M\"unzenberg}
\affiliation{Institut f\"ur Physik, Ernst-Moritz-Arndt Universit\"at Greifswald, 17489 Greifswald,
Germany}
\author{Luca~Perfetti}
\affiliation{Laboratoire des Solides Irradi\'es, CNRS UMR 7642 \& CEA-DSM-IRAMIS, Ecole
Polytechnique, F91128 Palaiseau, France}
\author{Tobias~Kampfrath}
\affiliation{Fritz Haber Institute of the Max Planck Society, 14195 Berlin, Germany}




\date{\today}

\begin{abstract}
Topological insulators constitute a new and fascinating class of matter with insulating bulk yet
metallic surfaces that host highly mobile charge carriers with spin-momentum locking. Remarkably,
the direction and magnitude of surface currents can be controlled with tailored light beams, but
the underlying mechanisms are not yet well understood. To directly resolve the \lq\lq birth" of
such photocurrents we need to boost the time resolution to the scale of elementary scattering
events ($\sim 10~\tn{fs}$). Here, we excite and measure photocurrents in the three-dimensional
model topological insulator $\tn{Bi}_2\tn{Se}_3$ with a time resolution as short as 20~fs by
sampling the concomitantly emitted broadband THz electromagnetic field from 1 to 40~THz.
Remarkably, the ultrafast surface current response is dominated by a charge transfer along the
Se-Bi bonds. In contrast, photon-helicity-dependent photocurrents are found to have orders of
magnitude smaller magnitude than expected from generation scenarios based on asymmetric
depopulation of the Dirac cone. Our findings are also of direct relevance for optoelectronic
devices based on topological-insulator surface currents.

\end{abstract}

\pacs{}

\maketitle

Many efforts in current solid-state research aim at pushing the speed of electronic devices from
the gigahertz to the terahertz ($1~\tn{THz}=10^{12}~\tn{Hz}$) range~\cite{Alamo2011} and at
extending their functionalities by the spin of the electron.\cite{Stamps2014} In these respects,
three-dimensional topological insulators (TIs) are a highly promising material class. While having
an insulating bulk, their surface is metallic due to a band inversion that is topologically
protected against external perturbations. $\tn{Bi}_2\tn{Se}_3$ is a model TI~\cite{zhang2009} as
its surface features a single pair of linear Dirac-type electron energy bands~\cite{qi2011} with
spin-velocity locking and forbidden $180\degree$ backscattering.\cite{moore2010} These properties
are ideal prerequisites to achieve large surface-current-induced spin polarizations.

Part of this large potential was demonstrated by recent works. They reported the exciting
possibility of launching TI surface currents by simply illuminating the sample with
light.\cite{mciver2012,duan2014,olbrich2014,luo2013,zhuShan2015,Tu2015,bas2015,kastl2015} The
direction of the photocurrent could be controlled through the polarization of the incident light
beam. The assignment to a surface process was bolstered by picosecond time-of-flight
measurements~\cite{kastl2015} showing that the ballistic photoinduced carriers were propagating at
a speed comparable to the band velocity of the Dirac states. There is, however, still an intense
debate about mechanisms leading to TI surface currents. Scenarios based on asymmetric depopulation
of the Dirac cone,\cite{mciver2012} transitions into other, higher-lying cones~\cite{kastl2015} and
asymmetric scattering of electrons~\cite{olbrich2014} have been proposed. To directly resolve the
generation of TI surface photocurrents, we need to boost the time resolution of the experiment from
so far $\sim 250$~fs and longer\cite{luo2013,zhuShan2015,Tu2015,bas2015,kastl2015} to the scale of
elementary scattering events, which can be shorter than 10~fs.

Here, we use ultrabroadband THz emission
spectroscopy~\cite{leitenstorfer1999,Leitenstorfer1999PRB,kampfrath2013} from 0.3 to 40~THz to
probe the ultrafast evolution of photocurrents in the model TI $\tn{Bi}_2\tn{Se}_3$ with
unprecedented time resolution. We identify distinct current sources: first, a slow drift of
photoinduced bulk charge carriers in the TI surface field. Second, for the first time, we observe a
new type of photocurrent, a surface shift current, which originates from an instantaneous
displacement of electron density along the Se-Bi bond. This current represents a dominant
charge-transfer excitation localized in a surface region of $\sim 3~\tn{nm}$ thickness, which is
the natural confinement scale of topological edge states. The modified electron density
redistributes with a time constant of 22~fs. Finally and remarkably, currents depending on the pump
helicity are found to have orders of magnitude smaller magnitude than expected from generation
scenarios based on asymmetric depopulation of the Dirac cone~\cite{mciver2012} (\fig{fig:3}d). This
result is not in contradiction to the previous observation of such currents in time-integrating
experiments.\cite{mciver2012} However, it shows that the generation of pump-helicity-dependent
photocurrents is surprisingly slow, thereby pointing to noninstantaneous
processes~\cite{olbrich2014} that are clearly distinct from the proposed instantaneous depopulation
scenario.


\begin{figure*}\centering\includegraphics[width=1\textwidth]{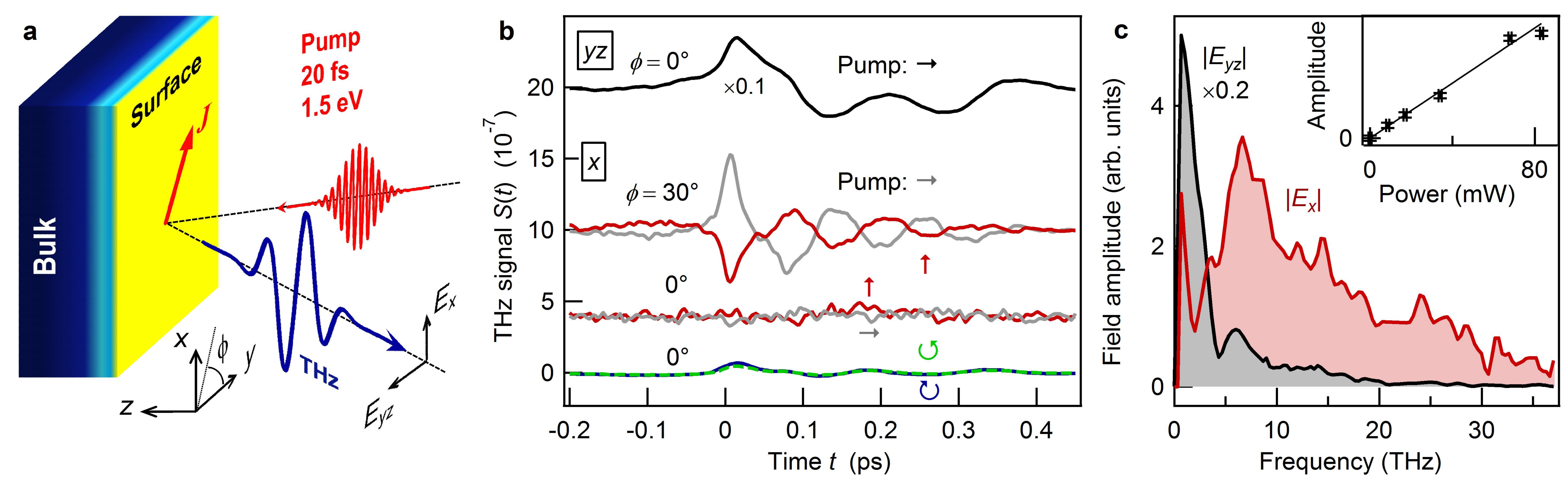}
\caption{Ultrafast photocurrent amperemeter: schematic and raw data. \textbf{a},~A
$\tn{Bi}_2\tn{Se}_3$ crystal is excited by a femtosecond laser pulse, resulting in a photocurrent
burst and, consequently, emission of a THz electromagnetic pulse. Measurement of the transient THz
electric field components $E_x(t)$ and $E_{yz}(t)$ by electrooptic sampling provides access to the
sheet current $\ve{J}(t)$ flowing inside the sample. \textbf{b},~Typical $x$- and $yz$-polarized
THz signals for various settings of pump polarization and sample azimuth~$\phi$. Signals are offset
for clarity. \textbf{c},~Amplitude spectra of the THz electric field directly behind the sample as
extracted from time-domain signals of panel~\textbf{b} (see text).}\label{fig:1}
\end{figure*}

\section*{Results}

\textbf{Ultrafast photocurrent amperemeter.} Our experimental setup is schematically depicted in
\fig{fig:1}a. A femtosecond laser pulse incident on the specimen launches a transient charge
current density $\ve{j}(z,t)$. This photocurrent, in turn, emits an electromagnetic pulse with
transient electric field $\ve{E}(t)$, in particular covering frequencies up to the THz range, as
expected from the inverse duration of the femtosecond stimulus. The measurement of $\ve{E}(t)$ over
a large bandwidth (0.3 to 40~THz) permits extraction of the sheet current density
\begin{equation}
\ve{J}(t)=\int\D z~\ve{j}(z,t)
\end{equation}
with ultrafast time resolution. More precisely, this approach allows us to separately determine the
current component $J_x$ directed along the $x$-axis and the component $J_{yz}$, which is a linear
combination of the Cartesian components $J_y$ and $J_z$ (\fig{fig:1}a). By virtue of a generalized
Ohm's law, the currents $J_x$ and $J_{yz}$  are, respectively, connected to the $x$-polarized
electric-field component $E_x$ and the perpendicular component $E_{yz}$ directly behind the sample
(\fig{fig:1}a). The THz near-fields $E_x$ and $E_{yz}$ are, respectively, obtained by measuring the
THz far-field using electrooptic sampling, resulting in the electrooptic signals $S_x$ and
$S_{yz}$, respectively (see Methods). THz waveforms are acquired for various settings of the pump
polarization and sample azimuth~$\phi$ (\fig{fig:1}a).

We use this approach to study a freshly cleaved, n-type $\tn{Bi}_2\tn{Se}_3$ single crystal in
ambient air~(see Methods). While photocurrents in the inversion-symmetric crystal bulk (space group
$\tn{D}^5_\tn{3d}$) cancel, optical excitation can in principle launch a current at the surface
(space group $\tn{C}_\tn{3v}$).\cite{liu2010} The surface region can be thought of as being
comprised of the air-crystal interface with locally relaxed lattice structure and simultaneously
hosting the Dirac surface states (thickness of $\sim 2~\tn{nm}$),\cite{zhang2010,Roy2014} followed
by a space-charge region with bent bulk bands (thickness of tens of
nanometers).\cite{mciver2012PRB,park2013}

In the following, we will show that our broadband current measurements allow us to discriminate
different types of photocurrents and their generation in the various surface regions. This goal is
achieved by first identifying two dominating components in the THz emission signal using symmetry
analysis. Based on the temporal structure of the two underlying photocurrent components, we can
finally assign these to microscopic generation scenarios.

\textbf{Raw data.} Typical THz electrooptic signal waveforms $S(t)$ from our $\tn{Bi}_2\tn{Se}_3$
sample are shown in \fig{fig:1}b. The THz waveforms depend sensitively on the setting of the THz
polarization ($x$ vs $yz$), the pump polarization and the sample azimuthal angle~$\phi$. The signal
amplitude grows linearly with increasing pump power, without any indication of saturation (inset of
\fig{fig:1}c). This behavior implies that the number of excited carriers is proportional to the
incident photon number.

\begin{figure}\centering\includegraphics[width=0.49\textwidth]{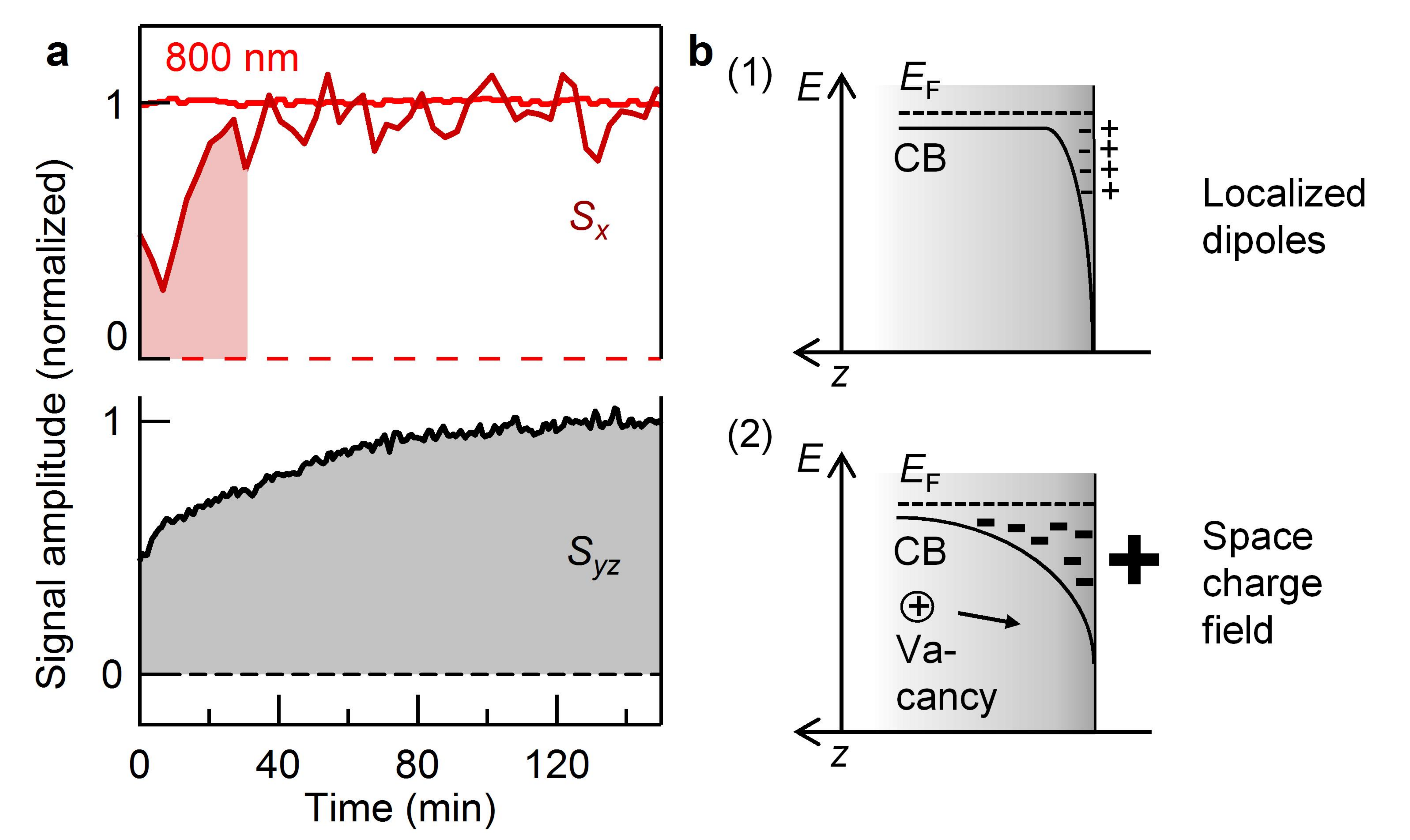}
\caption{
Impact of sample cleaving on THz emission. \textbf{a},~Evolution of the amplitudes of THz signals
$S_x$ and $S_{yz}$ and the reflectance of the 800~nm pump beam. The latter remains unchanged
showing that no sample damage occurs. \textbf{b},~Schematic of two symmetry-breaking processes
triggered by sample cleaving, (1)~formation of localized surface electric dipoles and (2)~band
bending due to migration of charged bulk Se vacancies toward the surface.}\label{fig:2}
\end{figure}
\begin{figure*}
\centering\includegraphics[width=1\textwidth]{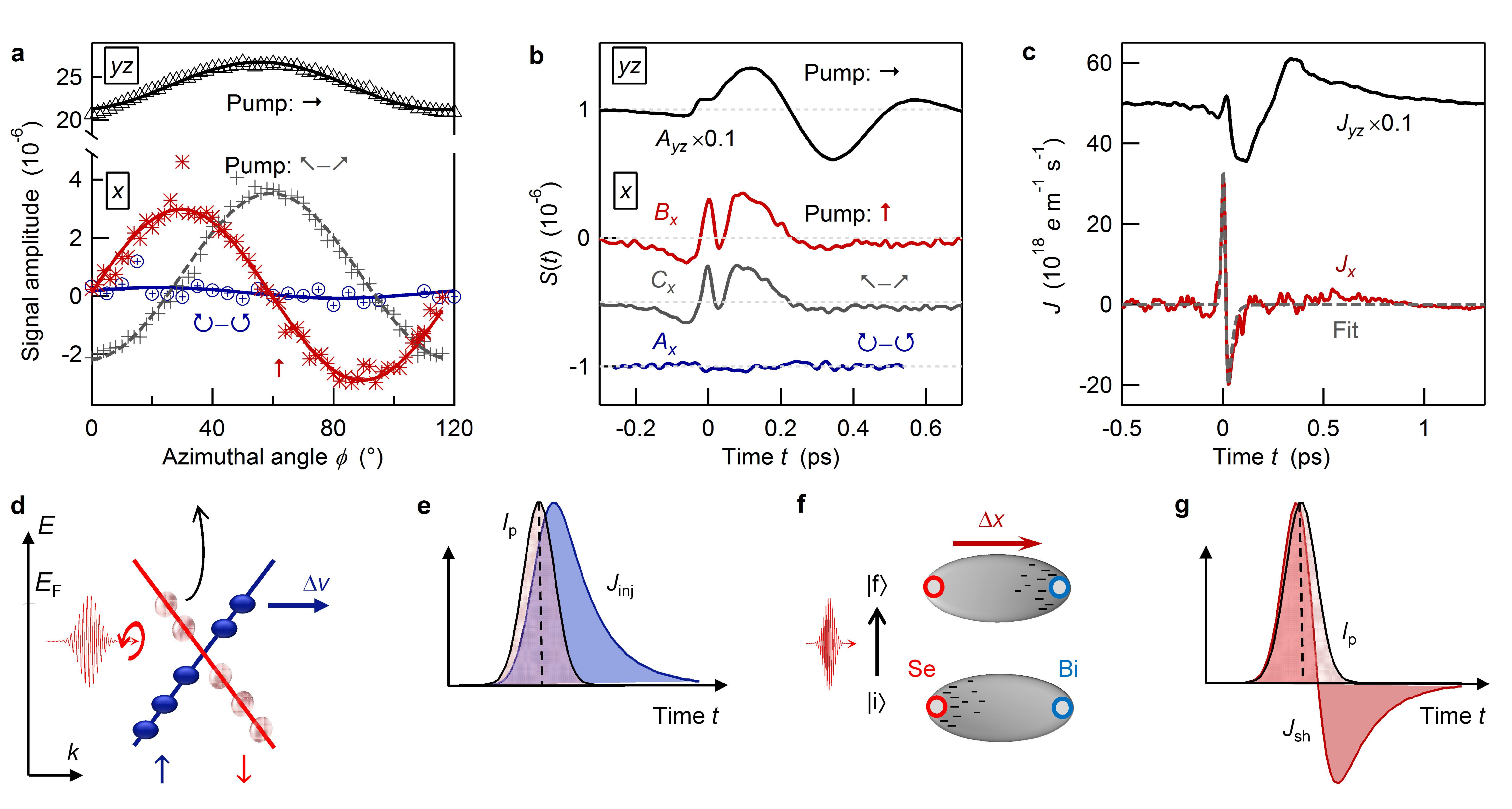} \caption{Extraction of the dominant
photocurrents and assignment. \textbf{a},~THz signal amplitude versus sample azimuth~$\phi$ for
various settings of the pump polarization ($\rightarrow$, $\uparrow$ and differential signals
$\nwarrow-\nearrow$ and $\circlearrowright-\circlearrowleft$). While $S_x$ exhibits a strong
$3\phi$-type-dependence, $S_{yz}$ is nearly independent of~$\phi$. \textbf{b},~Dominant temporal
components of signal sets $S_{yz}(t,\phi)$ and $S_x(t,\phi)$ for various pump polarizations,
extracted using \eq{eq:LinComb}. \textbf{c},~Source currents of the two dominant signal components.
The dynamics of these currents allow us to reveal the origin of the photocurrent.
\textbf{d},~Example of an injection-type photocurrent. The pump pulse promotes electrons from the
Dirac cone into other bands, thereby changing the electron band velocity. An asymmetric
depopulation of the Dirac cone and, thus, nonzero net current is achieved by using circularly
polarized light. \textbf{e},~Expected unipolar time-dependence of an injection-type photocurrent
$J_\tn{inj}(t)$ driven by a laser pulse with intensity envelope $I_\tn{p}(t)$. \textbf{f},~Scenario
of a shift photocurrent arising from an ultrafast transfer of electron density along the Se-Bi
bond. \textbf{g},~Resulting bipolar temporal shape of the sheet current density $J_\tn{sh}(t)$.
Signals in \textbf{b} and \textbf{c} are offset for clarity.}\label{fig:3}
\end{figure*}

As detailed in the following, we make the striking observation that the $x$- and $yz$-polarized
components of the emitted THz field (and, thus, $J_{x}$ and $J_{yz}$) behave very differently in
terms of their magnitude (\fig{fig:1}b), temporal shape (\fig{fig:1}b), behavior after sample
cleavage (\fig{fig:2}a) and azimuth-dependence (\fig{fig:3}a). First, as seen in \fig{fig:1}b,
$S_{yz}$ exhibits much larger amplitude than $S_x$ but evolves significantly more slowly. This
trend becomes even clearer when we apply an inversion procedure to these data to extract the THz
fields $E_{x}$ and $E_{yz}$ directly behind the sample (see Methods). The resulting spectral
amplitudes are displayed in \fig{fig:1}c as a function of angular frequency~$\omega$ and show that
$|E_x(\omega)|$ is much broader than $|E_{yz}(\omega)|$, indicating much faster temporal dynamics.

Second, to investigate the impact of surface modification on $S_{x}$ and $S_{yz}$, we freshly cleave
the sample and subsequently acquire THz signals continuously over 2~h with the sample exposed to
air. While the shape of the THz waveforms does not undergo measurable modifications, its global
amplitude increases by a factor of $\approx 2$ in the course of time~(\fig{fig:2}a). Note this rise
proceeds within 30~min for $S_x$ but significantly slower (within 100~min) for $S_{yz}$. We will
later relate this observation to distinct surface modification processes and use this information
to estimate the degree of surface localization of the currents $J_x$ and $J_{yz}$. In contrast to
$S_x$ and $S_{yz}$, measurable changes of the sample reflectance at a wavelength of 800~nm are not
observed, thereby ruling optical degradation of our sample out.

\textbf{Signal symmetries.} In addition to their different amplitude and temporal structure,
$S_{x}$ and $S_{yz}$ also depend very differently on the sample azimuth~$\phi$ (\fig{fig:1}a). To
quantify this behavior, we measure waveforms $S_{x}(t,\phi)$ and $S_{yz}(t,\phi)$ for an extended set
of $\phi$-values. To reliably extract an average signal amplitude for each $\phi$, we project the
time-domain signal on a suitable reference waveform (see Supplementary). The
resulting signal amplitude as a function of $\phi$ is displayed in \fig{fig:3}a. While $S_x$ is
almost fully modulated with a periodicity of $2\pi/3=120\degree$, $S_{yz}$ is dominated by a
constant offset.

The threefold rotational symmetry of the THz signals is consistent with the symmetry groups of
sample surface and bulk.\cite{zhang2009} Importantly, it allows us to significantly reduce the
large amount of experimental data contained in $S(t,\phi)$: for a given THz polarization ($x$ or
$yz$) and pump polarization, each two-dimensional set $S(t,\phi)$ can be written as a linear
combination of just three basis functions (see Supplementary),
\begin{equation}
S(t,\phi)=A(t)+B(t)\sin(3\phi)+C(t)\cos(3\phi).\label{eq:LinComb}
\end{equation}
Therefore, three basis signals $A(t)$, $B(t)$ and $C(t)$ fully characterize the entire data set
$S(t,\phi)$. They are, respectively, obtained by projecting $S(t,\phi)$ onto the mutually
orthogonal functions $1$, $\sin(3\phi)$ and $\cos(3\phi)$. Extracted waves are shown in
\fig{fig:3}b for the two THz polarizations and various pump polarizations.

We begin with considering the impact of the pump helicity on the photocurrent. The bottommost curve
in \fig{fig:3}b represents the $\phi$-independent component $A_x(t)$ of the difference of the
signals taken with right-handed ($\circlearrowright$) and left-handed ($\circlearrowleft$)
circularly polarized pump light. The amplitude of this waveform is comparable to the noise floor.
In other words, a helicity-dependent yet simultaneously $\phi$-independent THz signal is small and
below our detection threshold. This notion is consistent with time-domain raw data (blue vs green
trace in \fig{fig:1}b) and the absence of an offset in the $\phi$-dependence (blue curve of
\fig{fig:3}a).
We note that such small magnitude of the pump-helicity-dependent and $\phi$-independent
photocurrent does not contradict the previously reported observation of time-integrated
currents~\cite{mciver2012} as will be discussed further below.

\textbf{Photocurrents and assignment.} \figu{fig:3}b leads to another important conclusion of our
symmetry analysis: regardless of the pump polarization, all signals $S_x$ and $S_{yz}$ are,
respectively, dominated by just one fast and one slow waveform. We use these signals to extract the
underlying source currents (see Supplementary) which are displayed in \fig{fig:3}c. After an initial
onset, both ${J}_x$ and ${J}_{yz}$ change sign, indicating a backflow of charge. Note, however,
${J}_x$ proceeds on a much faster time scale than ${J}_{yz}$: the rise time from 10\% to 90\%
current maximum is $16~\tn{fs}$ for ${J}_x$ vs $120~\tn{fs}$ for ${J}_{yz}$. The respective decay
time constant are $27~\tn{fs}$ vs $200~\tn{fs}$.


To determine the origin of ${J}_x$ and ${J}_{yz}$ based on their ultrafast dynamics, we briefly
review known photocurrent generation
mechanisms.\cite{nastos2010,belinicher1980,lewis2014,glazov2014,olbrich2014,mciver2012,Ganichev2003,Ivchenko2008,Schmidt2015,Priyadarshi2012}
In general, optical excitation transfers electrons from initial states $\left|\tn{i}\right\rangle$
into final states $\left|\tn{f}\right\rangle$ [\fig{fig:3}(d)], followed by relaxation processes
such as scattering into other states, phonon emission and recombination.\cite{sobota2013}
Photocurrents can arise in both regimes, that is, during the optical transition and during the
subsequent relaxation. As our pump photon energy (1.57~eV) is much larger than the
$\tn{Bi}_2\tn{Se}_3$ band gap, numerous vertical interband transitions are
allowed~\cite{boschini2015} and expected to outnumber the contribution of phonon- or
impurity-assisted nonvertical transitions.\cite{Weber2008} To obtain a macroscopic net current,
inversion symmetry needs to be broken. In the relaxation regime, currents can arise from, for
instance, scattering by a noncentrosymmetric potential,\cite{belinicher1980,olbrich2014}
asymmetric recombination~\cite{glass1974} and carrier acceleration in an intrinsic surface field
(drift current).\cite{johnston2002,malevich2008}

As seen in \fig{fig:3}c, the slow current ${J}_{yz}(t)$ has a first peak (width of 120~fs) much
wider than the excitation pulse. Therefore, ${J}_{yz}$ cannot arise from the initial optical
transition. In fact, previous works on $\tn{Bi}_2\tn{Se}_3$ assigned the ${J}_{yz}$ component to a
carrier drift in the surface field, consistent with the strong dependence of ${J}_{yz}$ on the
doping level of $\tn{Bi}_2\tn{Se}_3$.\cite{luo2013,zhuShan2015,Tu2015} The double-peak structure of
${J}_{yz}$ indicates complex relaxation dynamics possibly involving strongly damped plasma
oscillations.\cite{malevich2008} These aspects are beyond the scope of this work, and we focus now
on the very fast, sub-100~fs photocurrent dynamics of ${J}_{x}$.

Along these lines, Sipe \emph{et al.}~\cite{Sipe2000,nastos2010} used perturbation theory to identify three
distinct mechanisms of photocurrent generation by an optical transition
$\left|\tn{i}\right\rangle\rightarrow\left|\tn{f}\right\rangle$: injection currents, shift currents
and optical rectification. \emph{Injection currents} ${J}_\tn{inj}$ arise because initial and final
state of the perturbed electron have different band velocity. An example is the asymmetric band
depopulation scenario~\cite{mciver2012} shown in \fig{fig:3}d: a circularly polarized pump excites
electrons from the Dirac cone into higher-lying states with different band slope (group velocity).
Therefore, for short enough excitation, ${J}_\tn{inj}$ should rise instantaneously to a magnitude
that scales with the average velocity change~$\Delta v$ and the density~$N$ of the excited
electrons. In this simplified model, the resulting current is
\begin{equation}
{J}_\tn{inj}=\sigma_\tn{inj}\Delta v~[\Theta(t)\tn{e}^{-t/\tau_\tn{inj}}]* I_\tn{p}\label{eq:inj}
\end{equation}
where the initial sheet charge density $\sigma_\tn{inj}=eN\Delta z_\tn{inj}$ is proportional to the
thickness $\Delta z_\tn{inj}$ of the emitting sheet. Furthermore, $\Theta(t)$ is the unit step
function, and the exponential accounts for relaxation of the current with time constant
$\tau_\tn{inj}$. Backflow of electrons is diffusive~\cite{apostolopoulos2014} and ignored on the
short timescales considered here. Finally, the convolution with the pump intensity envelope
$I_\tn{p}(t)$ (normalized to unity) accounts for the shape of the pump pulse.

\emph{Shift currents}\cite{nastos2006}, on the other hand, arise when the electron density
distribution of the excited state $\left|\tn{f}\right\rangle$ is spatially shifted with respect to
$\left|\tn{i}\right\rangle$ (\fig{fig:3}f). For short excitation, this process leads to a step-like
charge displacement $\Delta x_\tn{sh}\Theta(t)$ whose temporal derivative is proportional to the
shift current ${J}_\tn{sh}$. With arguments analogous to the injection case, we obtain
\begin{equation}
{J}_\tn{sh}=\sigma_\tn{sh}\Delta x~\frac{\partial}{\partial t} [\Theta(t)\tn{e}^{-t/\tau_\tn{sh}}]* I_\tn{p}, \label{sh}
\end{equation}
with $\sigma_\tn{sh}=eN\Delta z_\tn{sh}$. This model implies $J_\tn{sh}$ initially follows the
profile of $I_\tn{p}(t)$ and becomes bipolar if the relaxation time $\tau_\tn{sh}$ is comparable to
or longer than the pump duration (\fig{fig:3}g). Finally, \emph{optical rectification} can be
understood as a nonresonantly driven virtual charge displacement. This effect is typically two
orders of magnitude smaller than resonant optical transitions~\cite{cote2002} and will not be
considered further.

Note the characteristic shape of the ultrafast currents ${J}_\tn{inj}$ and ${J}_\tn{sh}$ is very
distinct: unipolar (\fig{fig:3}e) and bipolar asymmetric (\fig{fig:3}g). Having understood how the
temporal shape of a current is intrinsically linked to its origin, we now look for such
fingerprints in our data (\fig{fig:3}c). Indeed, we find that the measured photocurrent $J_x$
(\fig{fig:3}c) has bipolar asymmetric temporal shape, the unambiguous fingerprint of a shift
current. In addition, fitting \eq{sh} to ${J}_x$ yields excellent agreement (\fig{fig:3}c) for a
pump duration of 23~fs, $\tau_\tn{sh}=22~\tn{fs}$ and $\Delta x_\tn{sh}\Delta z_\tn{sh}\approx
36~\tn{\AA}^2$. In this procedure, we use the excitation density ($N=6.9\times10^{24}~\tn m^{-3}$)
as inferred from the absorbed pump fluence ($4~\mu\tn{J cm}^{-2}$), the pump photon energy
(1.57~eV) and the pump penetration depth (24~nm at $1/\tn{e}$ intensity).\cite{mciver2012PRB} The
thickness $\Delta z_\tn{sh}$ of the shift-current sheet will be determined next.

\textbf{Surface localization.} Our photocurrent measurements directly reveal an ultrafast shift
current and a drift current in the time domain. It is so far, however, unclear to which extent
these currents are localized at the surface. Although second-order optical probes such as THz
emission and sum-frequency generation are only operative at the surface~\cite{shen1984} of samples
with inversion-symmetric bulk, the region of broken inversion symmetry extends over a certain
depth. For example, the bulk drift current dominating ${J}_{yz}$ (\fig{fig:3}c) is known to flow in
a layer whose thickness is given by the Thomas-Fermi screening length of the surface depletion
field, which is $\sim 16~\tn{nm}$ in our sample.\cite{mciver2012PRB}

Proving surface sensitivity is a well-known issue of non-linear optics and has a common
solution~\cite{shen1984,zhuShan2015}: modify the sample surface and monitor the impact on the
signal. We modify the $\tn{Bi}_2\tn{Se}_3$ surface by cleaving, which is known to trigger transport
processes on a time scale of 100~min.\cite{zhuShan2015,mciver2012} Extensive studies on such
$\tn{Bi}_2\tn{Se}_3$ aging effects~\cite{xu2014,mciver2012PRB,park2013,benia2011} revealed two
mechanisms (see schematic of \fig{fig:2}b): (1)~formation of an electric-dipole layer due to charge
transfer from adsorbates~\cite{benia2011,park2013} or surface lattice relaxation~\cite{Roy2014} and
(2)~migration of bulk defects, mainly Se vacancies, toward the surface.\cite{park2013,xu2014} While
the surface dipole layer is very localized (thickness of $\sim 3~\tn{nm}$),\cite{park2013} the
redistribution of bulk Se vacancies induces a more extended space-charge region (thickness of tens
of nanometers).~\cite{park2013,xu2014}

We now relate these processes to the amplitude evolution of ${J}_{yz}$ and ${J}_x$ following sample
cleaving (\fig{fig:2}a). The drift current amplitude ${J}_{yz}$, intrinsically linked to the
strength of the space-charge field, follows the redistribution of Se vacancies within 100~min. In
contrast, ${J}_x$ rises much faster (30~min), thereby showing that ${J}_x$ originates from a layer
in which a competing process dominates the signal modification. Consequently, we assign the fast
time scale of ${J}_x$ to the formation of the localized surface dipoles, suggesting ${J}_x$ flows
in a surface layer with a thickness $\Delta z_\tn{sh}$ of less than 3~nm. This value and the above
extracted estimate for $\Delta x_\tn{sh}\Delta z_\tn{sh}$ imply the shift distance $\Delta
x_\tn{sh}$ is on the order of 1~\AA.

\section*{Discussion}

Summarizing our results, we have shown that our ultrabroadband THz emission data are fully
consistent with the notion that (i)~the photocurrent ${J}_x$ arises from an instantaneous
photoinduced shift of charge density by $\sim 1~\tn{\AA}$ in a $\sim 3~\tn{nm}$ thick surface
region of $\tn{Bi}_2\tn{Se}_3$. The displacement relaxes on a very fast time scale of 22\,fs. The
much slower current ${J}_{yz}$ is dominated by a drift current of optically excited carriers in the
surface field. (ii)~A helicity-dependent and simultaneously azimuth-independent photocurrent is
smaller than our detection threshold of ${10^{18}~e\,\tn{m}^{-1}\,\tn{s}^{-1}}$. This assertion is
also valid for other injection-type transport scenarios such as photon-drag
currents.\cite{glazov2014} It is instructive to discuss these observations and compare them to
previous works.

\textbf{Surface shift currents.} Finding~(i) represents the first observation of a surface shift
current, which was theoretically predicted by Cabellos \emph{et al.} very
recently.\cite{cabellos2011} We emphasize that revealing the time-domain fingerprint of shift
currents relies on the 20~fs time resolution of our experiment. Longer pump pulses can easily
obscure this signature, even in materials with broken bulk inversion symmetry.\cite{laman2005} The
threefold azimuthal symmetry of $J_x$ (\fig{fig:3}a) suggests the electron density is displaced
along the $120\degree$-ordered p-type Se-Bi bonds.\cite{mishra1997} As Bi and Se atoms lie in
different layers, the shift current also has a $z$-component with a strength comparable to ${J}_x$,
consistent with the sharp peak present in ${J}_{yz}$ at $-10~\tn{fs}$ [\fig{fig:3}(c)].

Our results show that the displacement of bound charges occurs in a sheet with thickness $\Delta
z_\tn{sh}\sim 3~\tn{nm}$. This notion is consistent with reports~\cite{Roy2014} showing that only
the first quintuple layer exhibits inversion asymmetry on the order of 10\%. This fact indicates
that the current is generated in a layer where the Dirac states are expected to dominate
transport.\cite{zhang2010} The shift distance $\Delta x_\tn{sh}\sim1~\tn{\AA}$ compares well to
reported charge shifts on the order of the bond length ($\sim 3~\tn{\AA}$) in noncentrosymmetric
semiconductors.\cite{nastos2006} In addition, the electron density associated with the Dirac states
is known to shift from Se toward Bi atoms when energies below and above the the Dirac point are
considered.\cite{zhang2009,mishra1997} The charge shift relaxes within 22~fs which coincides with
the time scale known for depopulation of the optically populated antibonding
states.\cite{sobota2013} Therefore, the assignment of $J_x$ to a surface shift current is fully
compatible with previous works.

\textbf{Helicity-dependent currents.} Result~(ii), the absence of a circular photocurrent, is
surprising and imposes significant constraints on the generation mechanism and shape of this
current. McIver \emph{et al.} observed a helicity-dependent time-integrated photocurrent and
suggested it to arise from asymmetric depopulation of the Dirac cone by optical transitions into
rapidly decaying bulk states (\fig{fig:3}f). Based on this injection-type scenario, we use
\eq{eq:inj} to estimate the initial ballistic sheet-current density as $Nev_\tn{D} \Delta
z_\tn{D}$, where $v_\tn{D}=0.5~\tn{nm\,fs}^{-1}$ is the band velocity in the Dirac
cone,\cite{sobota2013} $\Delta z_\tn{D}=2$\,nm the \lq\lq thickness" of the Dirac
states,\cite{zhang2010} and $N$ is the excitation density. The resulting magnitude of
$10^{22}~e\,\tn{m}^{-1}\,\tn{s}^{-1}$ is four orders of magnitude larger than the maximum current
measured in our experiment. Therefore, our measurements render the simple asymmetric-depopulation
scenario very unlikely.

This result is supported by comparing the magnitudes of the helicity-dependent photocurrent seen in
the time-integrated~\cite{mciver2012} and in our time-resolved measurements. Assuming the transient
current has rectangular temporal shape with amplitude $J_0$ and duration $\tau_0$, a
time-integrated measurement~\cite{mciver2012} yields an average sheet-current density of
$\overline{J}=J_0 \tau_0 f_\tn{rep}$, where $f_\tn{rep}\sim 100$~MHz is the repetition rate of
commonly used femtosecond laser oscillators. For small $\overline{J}$ and long $\tau_0$, $J_0$ may
drop below our detection threshold of ${10^{18}~e\,\tn{m}^{-1}\,\tn{s}^{-1}}$. Using the value
$\overline{J}\sim 10^{13}~e\,\tn{m}^{-1}\,\tn{s}^{-1}$ obtained in~\Ref{mciver2012} under
excitation conditions similar to ours, we find the helicity-dependent current must flow for a
duration $\tau_0>100$~fs to be below our detection threshold. Such a relatively long current
lifetime is not in favor of the asymmetric depopulation scenario (\fig{fig:3}f)\cite{mciver2012}
since the asymmetry of photogenerated holes in the Dirac cone is known to decay on a 40~fs time
scale.\cite{Park2010}
Therefore, our observations point to indirect and slower generation mechanisms of
pump-helicity-dependent photocurrents, for example asymmetric electron scattering, as proposed for
near-equilibrium electrons.\cite{olbrich2014}

In conclusion, we have measured the dynamics of ultrafast photocurrents on the surface of the
three-dimensional model TI $\tn{Bi}_2\tn{Se}_3$ with a time resolution as short as 20~fs. We find
that the peak amplitude of pump-helicity-dependent photocurrents is much smaller than predicted
based on previous models. Its duration is inferred to exceed 100~fs. These results point to
noninstantaneous generation mechanisms of the pump-helicity-dependent photocurrent and call for
improved models and theories. In addition, we have for the first time observed a surface shift
photocurrent which arises from a charge displacement on the TI surface. This current is potentially
interesting for ultrafast optical manipulation of the TI surface, ultimately thereby modifying its
topological properties.\cite{liu2015TIswitch} Finally, our results highlight broadband THz emission
spectroscopy as a novel and highly sensitive probe of surfaces.



\section*{Methods}

\textbf{Sample details.} Single crystals of Ca-doped $\tn{Bi}_2\tn{Se}_3$ were grown by the
Bridgman-Stockbarger method by pulling a sealed quartz ampoule in a vertical temperature gradient.
Hall measurements\cite{hruban2011} yield a bulk hole density of $1.34\times10^{17}~\tn{cm}^{-3}$
and a mobility of $275~\tn{cm}^2\,\tn{V}^{-1}\,\tn{s}^{-1}$. From angle-resolved photoelectron
spectroscopy (ARPES), we extract a conduction-band electron mass of 0.115 bare electron
masses.\cite{Xia2009} A fresh surface is obtained by cleaving using adhesive tape. After exposition
to air, ARPES measurements confirm the presence of Dirac surface states with the Fermi energy
located 160~meV above the bulk conduction band minimum.

\textbf{Ultrafast amperemeter.} Laser pulses (duration of $\approx 20~\tn{fs}$, center wavelength
of 800~nm, energy 1~nJ) from a Ti:sapphire oscillator (repetition rate 80~MHz) are focused onto the
sample (beam diameter of $200~\mu\tn{m}$ full-width at half intensity maximum) under $45\degree$
angle of incidence, resulting in an average intensity $<0.3\,\tn{kW}\,\tn{cm}^{-2}$, well below
sample damage threshold. The specularly emitted THz pulse is focused onto an electrooptic crystal
in which the THz electric field is detected by broadband electrooptic sampling.\cite{ferguson2002}
We use a (110)-oriented GaP crystal (thickness of $250~\mu\tn{m}$) owing to its relatively flat and
broadband response function.\cite{kampfrath2013} The only exception are the measurements of the
two-dimensional data set $S(t,\phi)$ (\figs{fig:3}a and \ref{fig:3}b) which are sped up by using
(110)-oriented ZnTe crystal (thickness of $300~\mu\tn{m}$) having an enhanced detector response at
the expense of reduced bandwidth.

To proceed from the measured electrooptic signal $S(t)$ to the THz electric field $\ve{E}(t)$
directly behind the sample, we also measure the transfer function of our spectrometer (see
Supplementary). We finally obtain the source current $\ve{J}(t)$ by employing a
generalized Ohm's law (see Supplementary). Optical wave plates are used to set the
polarization state of the pump pulse to circular or linear with arbitrary rotation angle. A THz
wire-grid polarizer (field extinction ratio of $10^{-2}$) allows us to measure the $x$- and
$yz$-components $E_{x}$ and $E_{yz}$ of the THz electric field separately, thereby disentangling
current components $J_x$ and $J_{yz}$, the latter being a linear combination of $J_y$ and $J_z$
(\fig{fig:1}a). To ensure the electrooptic THz detector has an identical response to $E_x$ and
$E_{yz}$, a wire-grid polarizer with $45\degree$ orientation is placed in front of it.

\section*{Acknowledgments}

We thank S.D. Ganichev, A.D. Bristow and N.P. Armitage for stimulating and fruitful discussions. We
are grateful to the German Science Foundation (DFG) for financial support through priority program
SPP 1666 \lq\lq Topological insulators: materials, fundamental properties, devices" (grant no.
KA~3305/3-1). L.P. and M.K. thank for support by grant ANR-13-IS04-0001-01.

\bibliography{PRL_shiftCurrent_Vol30}

\end{document}